\documentclass[referee]{jfm}
\usepackage{graphicx}
\usepackage{natbib}
\usepackage{amssymb}

\ifCUPmtlplainloaded \else
  \checkfont{eurm10}
  \iffontfound
    \IfFileExists{upmath.sty}
      {\typeout{^^JFound AMS Euler Roman fonts on the system,
                   using the 'upmath' package.^^J}%
       \usepackage{upmath}}
      {\typeout{^^JFound AMS Euler Roman fonts on the system, but you
                   don't seem to have the}%
       \typeout{'upmath' package installed. jfm.cls can take advantage
                 of these fonts,^^Jif you use 'upmath' package.^^J}%
       \providecommand\mathup[1]{##1}%
       \providecommand\mathbup[1]{##1}%
      }
  \else
    \providecommand\mathup[1]{#1}%
    \providecommand\mathbup[1]{#1}%
  \fi
\fi

\ifCUPmtlplainloaded \else
  \IfFileExists{amsbsy.sty}
    {\typeout{^^JFound the 'amsbsy' package on the system, using it.^^J}%
     \usepackage{amsbsy}}
    {\providecommand\boldsymbol[1]{\mbox{\boldmath $##1$}}}
\fi

\newcommand\etal{\mbox{\textit{et al.}}}

\title[Dynamics of zero-P convection]{Dynamics of zero-Prandtl number convection near the onset}

\author[P. Pal, S. Paul, P. Wahi and M. K. Verma]{P.\ns P\ls A\ls L$^1$,\ns S.\ns P\ls A\ls U\ls L$^2$,\ns P.\ns W\ls A\ls H\ls I$^3$ \and M.\ns K.\ns V\ls E\ls R\ls M\ls A$^2$}

\affiliation{$^1$Department of Mathematics, National Institute of Technology, Durgapur-713~209, India\\[\affilskip]
$^2$Department of Physics, Indian Institute of Technology, Kanpur-208~016, India\\[\affilskip]
$^3$Department of Mechanical Engineering, Indian Institute of Technology, Kanpur-208~016, India}

\begin{document}

\label{firstpage}
\maketitle

\begin{abstract}
In this paper we present various convective states of zero-Prandtl number Rayleigh-B\'enard convection using direct numerical simulations (DNS) and a 27-mode low-\break dimensional model containing the energetic modes of DNS.  The origin of these convective states have been explained using bifurcation analysis.  The system is chaotic at the onset itself with three coexisting chaotic attractors that are born at two codimension-2 bifurcation points.  One of the bifurcation points with  a single zero eigenvalue and a complex pair $(0,\pm i \omega)$ generates chaotic attractors and associated periodic, quasiperiodic, and phase-locked states that are related to the wavy rolls observed in experiments and simulations.  The frequency of the wavy rolls are in general agreement with  $\omega$ of the above eigenvalue of the stability matrix.   The other bifurcation point with a double zero eigenvalue produces the other set of chaotic attractors and ordered states such as squares, asymmetric squares, oscillating asymmetric squares, relaxation oscillations with intermediate squares, some of which are common to the 13-mode model of  \cite{Pal:EPL_2009}.
\end{abstract}


\section{Introduction}

Origin of instabilities, patterns, and chaos in Rayleigh-B\'enard convection (RBC) is an interesting problem in fluid dynamics~\cite[]{chandra:book,busse:TAP_1981,jkb:book,manneville:book, ahlers:RMP_2009}.   Prandtl number, $P$ (ratio of kinematic viscosity $\nu$ and thermal diffusivity $\kappa$) and Rayleigh number, $R$ (ratio of buoyancy and dissipative terms) are the two critical parameters for RBC.   Some of the key and difficult problem in this field are related to the onset of convection for zero-Prandtl number (zero-P) and low-Prandtl number (low-P) fluids.  For zero-P and low-P convection, the nonlinear term of the Navier-Stokes equation plays an important role, and it generates vertical vorticity resulting in three-dimensional patterns and the associated secondary oscillatory patterns~\cite[]{busse:JFM_1972, clever:JFM_1974,  busse:jfm_1984}.  In the present paper we will study various patterns and chaos for zero-Prandtl number convection using bifurcation analysis.  We use direct numerical simulations and a low-dimensional model for this purpose.

Low-Prandtl number fluids, for example, mercury ($P \approx 0.02$), liquid sodium ($P \approx 0.01$), solar plasma in the convective zone ($P \sim 10^{-8}$),  exhibit interesting convective patterns and chaos \cite[]{croquette:CP_1989_a, croquette:CP_1989_b,cross:RMP_1993, bodenschatz:ARFM_2000, das:PRE_2000, stein:solar}.  In terrestrial experiments, the Prandtl number cannot go below that of liquid sodium ($\sim 0.01$).   In addition, the visualization of flow patterns inside low-P fluids like mercury and sodium is quite difficult.  Due to these limitations of experiments, numerical simulations are very significant  in the study of low-P and zero-P convection.  The numerical work of \cite{thual:zeroP_1992} indicate that the properties of low-P convection as $P \rightarrow 0$ are quite close to that of zero-P convection.  Hence low-P convection as $P \rightarrow 0$ appears to approach zero-P convection as a limiting case.  Therefore,  zero-P convection is very useful for understanding the properties of low-P convection.   Even though numerical analysis of zero-P convection is quite tricky due to its inherent instabilities near the onset, it provides certain advantages.   For $P=0$  the thermal modes are slaved to the velocity modes, hence the number of independent variables are less than those required for low-P analysis.  Also, the time steps for numerical simulations of very low-P convection could be very small due to the stiffness of the equations, which is not a limitation for zero-P convection \cite[]{thual:zeroP_1992}.  These features led us to investigate zero-P convection in detail for understanding the convective patterns and chaos in low-P fluids.

\cite{thual:zeroP_1992} was one of the first to simulate zero-P convection for free-slip and no-slip boundary conditions.  He reported various supercritical oscillatory instabilities, regular and chaotic patterns, etc.\ in his simulations.   The patterns observed by Thual are  two-dimensional rolls, periodic and quasiperiodic rolls, squares, travelling waves, etc.   For low-P fluids, \cite{meneguzzi:jfm_1987} performed numerical simulations for $P = 0.2$ under stress-free boundary conditions, and for  $P = 0.025$ under no-slip boundary conditions, and observed similar patterns. \cite{clever:pf_1990} found travelling wave solutions for low-P convection under no-slip boundary conditions.  These observations indicate that zero-P convection contains pertinent features of low-P convection.  Also, free-slip and no-slip boundary conditions exhibit similar convective flow patterns.

The oscillatory instabilities, and regular and chaotic patterns found in numerical simulations have also have been observed in convection experiments on mercury,  air, and liquid sodium  \cite[]{Rossby:lowP, krishnamurti:JFM_1970, krishnamurti:JFM_1973, maurer:JPL_1980,libchaber:JPL_1982, libchaber:physica_1983}.  Some of the most commonly observed patterns in experiments are stationary, periodic, quasiperiodic, chaotic, and travelling rolls, as well as squares, asymmetric squares, and phase locked states.  Chaotic rolls have been observed to appear through both period doubling and quasiperiodic routes in some of these experiments.  

Convection experiments of \cite{willis:pf_1967}, \cite{krishnamurti:JFM_1970, krishnamurti:JFM_1973},  \cite{busse:JFM_1974}, and numerical simulations of \cite{lipps:jfm_1976}, \cite{bolton:JFM_1985}, \cite{clever:pf_1990}, \cite{meneguzzi:jfm_1987}, \cite{thual:zeroP_1992}, \cite{ozoe:NHT_1995} indicate the presence of oscillatory instability of the two-dimensional convective rolls and the resultant stable wavy rolls.  Busse and coworkers~\cite[]{busse:JFM_1972, clever:JFM_1974,  busse:jfm_1984} showed using perturbative analysis that the two-dimensional rolls become unstable to oscillatory three-dimensional disturbances when the amplitude of the convective motion exceeds a finite critical value.   According to \cite{busse:JFM_1972}, the condition for the instability takes the form
\begin{equation}
\frac{R_t}{R_c} -1 \approx 0.310 P^2
\end{equation}
where $R_c$ is the critical Rayleigh number for the onset of convection, and $R_t$ is the Rayleigh number where oscillatory instability starts.  In addition, the time period of the oscillations measured in the units of $d^2/\nu$ (viscous time scale) is independent of $P$ \cite[Eq.~(5.2) of][]{busse:JFM_1972}.  In a related work, \cite{fauve:1987} investigated the origin of instabilities in low-P convection using phase dynamical equations and argued that the instability always saturates into travelling waves. In the present paper we will explore the origin of oscillatory instability using numerical simulations and bifurcation theory.

Origin of plethora of convective patterns observed in convective experiments and numerical simulations can be quite intricate.  Each run of a convective simulation takes significantly long time, so it is not possible to scan the parameter space minutely for deciphering detailed bifurcation scenario.  Experiments too have their complexities and limitations.  The large number of modes present in simulations and experiments tend to obscure the underlying dynamics.  These difficulties are circumvented by a  powerful and complimentary approach in which the system is analyzed using an appropriately constructed low-dimensional model.   Relatively low computational cost for running low-dimensional models, and the ease of construction of the bifurcation diagrams are some of the distinct advantages of these models compared to the experiments and simulations.  However care must be taken to take all the relevant modes of the system for constructing the low-dimensional models.  

Number of active modes near the onset of convection is not very large, so low-dimensional models consisting of these active modes are very useful for analyzing this regime.    \cite{kumar:JP_1996}  showed using a 6-mode model of zero-P convection that the growth of the 2D rolls saturate through the generation of vertical vorticity (wavy nature). \cite{kumar:burst_2006} observed critical bursting in the above model during the saturation. \cite{Pal:PRE_2002} explained the mechanism of selection of the square patterns using a 15-mode model of the zero-P RBC.   Recently \cite{Pal:EPL_2009} constructed a 13-mode model using the energetic modes of DNS, and performed a bifurcation analysis using the model and simulation results.  Using the bifurcation diagram, \cite{Pal:EPL_2009}  could explain  the origin of squares (SQ), asymmetric squares (ASQ), oscillating asymmetric squares (OASQ), relaxation oscillations with an intermediate square regime (SQOR), and three kinds of chaotic attractors.  In addition, the above patterns were observed in both DNS and models.  Earlier \cite{thual:zeroP_1992} had observed SQ and SQOR patterns in his simulations.  \cite{jenkins:1984} studied the transition from 2D rolls to square patterns in RBC for different Prandtl numbers using analytical tools. \cite{knobloch:jp_1992}  studied  the stability of the SQ patterns using a complementary procedure called the amplitude equations.

A major limitation of the 13-mode model of \cite{Pal:EPL_2009} was the absence of wavy rolls near the onset.  This limitation is overcome by extending this model to a 27-mode model by incorporating the corresponding dominant modes.  In the present paper we perform a bifurcation analysis of this model and explore the origin of the various convective patterns of zero-P convection with special emphasis on the oscillatory instability and related wavy roll patterns.  All the features of the 13-mode model are reproduced in the 27-mode model by construction.  We will show in our discussion that properties of the wavy rolls of the  27-mode model matches reasonably well with those observed in experiments and simulations.  

The organization of the paper is as follows: In section \ref{sec:hydrosystem}, we describe the basic hydrodynamic system considered for the study. The low dimensional model will be derived in section \ref{sec:lowdmodel}.  Section \ref{sec:model_results} contains the results of the numerical simulations and the low-dimensional model.  Various bifurcation diagrams are described in this section. A brief study of the wavy rolls observed in the low-dimensional model is presented in section \ref{sec:wavy_rolls}. We finally conclude in section \ref{conclusion}.

%%%%%%%%%%%%%%%%%%%%%%%%%%%%%%%%%%%
\section{Governing equations and direct numerical simulations}\label{sec:hydrosystem}

The RBC system consists of a conducting fluid of kinematic viscosity
$\nu$, thermal diffusivity $\kappa$, and coefficient of volume
expansion $\alpha$ confined between two conducting plates separated
by a distance $d$ and heated from below. In the zero-P limit the
equations under Boussinesq approximation take the
form~\cite[]{spiegel,thual:zeroP_1992}
\begin{eqnarray}
\partial_t(\nabla^2 v_3) &=& \nabla^4 v_3 + R \nabla^2_H \theta - \hat{\bf e}_3\cdot\nabla\times
           \left[(\mbox{\boldmath $\omega$}{\cdot}\nabla){\bf v}
           -( {\bf v}{\cdot}\nabla)\mbox{\boldmath $\omega$}  \right],
\label{eq:v3}\\
\partial_t \omega_3 &=& \nabla^2 \omega_3
          +\left[(\mbox{\boldmath $\omega$}{\cdot}\nabla) v_3
          -({\bf v}{\cdot}\nabla)\omega_3\right],
\label{eq:omega3}\\
  {\nabla}^2 \theta&=& - v_3,\label{eq:theta}\\
\nabla{\cdot}{\bf v} &=& 0, \label{eq:continuity}
\end{eqnarray}
where ${\bf v}(x,y,z,t) \equiv (v_1,v_2,v_3)$ is the velocity field,
$\theta(x,y,z,t)$ is the  deviation in the temperature field from the
steady conduction profile, $\mbox{\boldmath $\omega$} \equiv
(\omega_1, \omega_2, \omega_3) \equiv \nabla\times{\bf v}$ is the
vorticity field, $\hat{\bf e}_3 $ is the vertically
directed unit vector, and $ \nabla_H^2 = \partial_{xx} +
\partial_{yy} $  is the horizontal Laplacian. The equations have
been nondimensionalized using  $d$ as the length scale, $d^2/\nu$ as
the time scale, and $\nu \beta d/\kappa$ as the temperature scale,
where $\beta$ is the uniform temperature gradient.   The two
nondimensional parameters in the equations are the Rayleigh number $R =
\alpha \beta g d^4/{\nu\kappa}$ and the Prandtl number $P = \nu/\kappa$,
where $g$ is the acceleration due to gravity.    In the following
discussions we will also use the reduced Rayleigh number $r =
R/R_{c}$ as a parameter.

We consider perfectly conducting boundary conditions for the top and
bottom plates along with the free-slip boundary condition for the
velocity field.  Consequently
\begin{equation}
v_3 = \partial_{3}v_1 =  \partial_{3}v_2  = \theta = 0 ,\quad
\mbox{at}\quad z = 0, 1. \label{eq:bc}
\end{equation}
We assume periodic boundary conditions along the horizontal directions.

Equations~(\ref{eq:v3}-\ref{eq:continuity}) are numerically solved
using direct numerical simulations (DNS) under the above boundary conditions (Eqs.~\ref{eq:bc}).  DNS were performed using a pseudo-spectral code TARANG~\cite[]{canuto,tarang} in a box with aspect ratio $\Gamma_x = 2\sqrt{2}, \Gamma_y = 2\sqrt{2}$. Various grid resolutions, $32\times 32\times 32$, $64\times 64\times 64$, have been used for the simulations.   These grids provide well-resolved simulations near the onset of convection.  We used fourth-order Runge-Kutta scheme (RK4) for time advancement. Each run was carried out till the system reaches a steady state. The DNS runs were performed for $ 0.98 \le r \le 1.25$.

We perform around 200 simulation runs, yet it is insufficient to construct the bifurcation diagram of the system in detail.   For this purpose we construct a low-dimensional model, which will be described in the next section.

\section{Low-dimensional model}\label{sec:lowdmodel}

\begin{figure}
\begin{center}
\includegraphics[height=!,width=12cm]{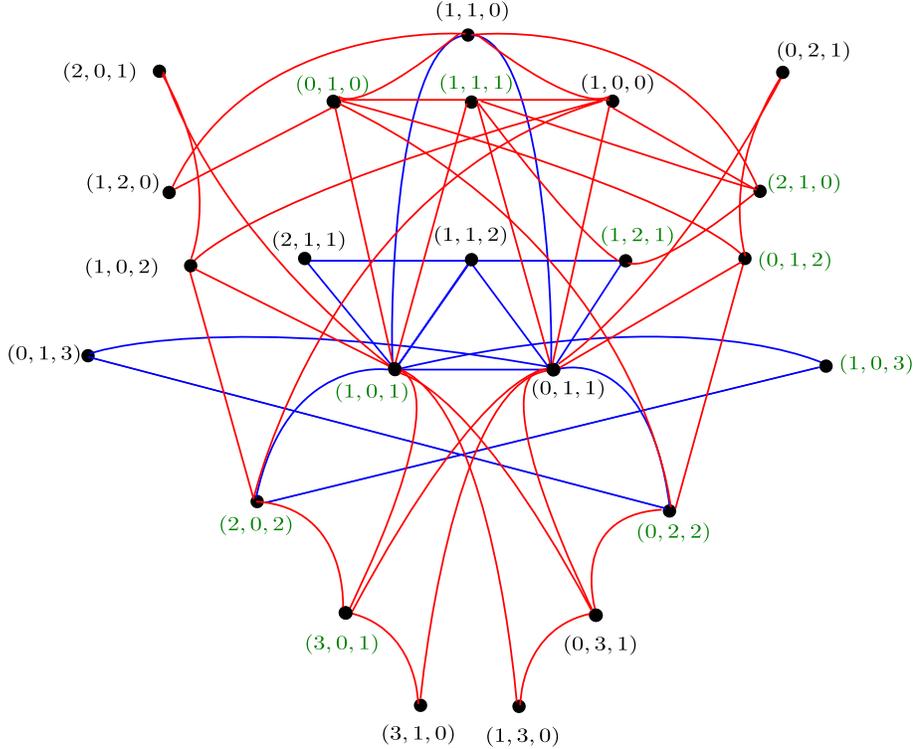}
\end{center}
\caption{The interacting modes of the 27-mode model. Interactions corresponding to the 13-mode model are shown in blue, whereas new interactions unique to 27-mode model are shown in red. The modes in green are active in the wavy roll flow patterns.} 
\label{fig:modes_lowdim}
\end{figure}

 We identify 27 Fourier modes that have  significant energy (approximately 1\% or more of the total energy) in the DNS runs and  create a low-dimensional model.  The modes of the model are shown in Fig.~\ref{fig:modes_lowdim} and they account for approximately 98\% of the total energy.  The triangles represent  the interacting  triads. Note that only some of the interacting triads of the model have been shown in the figure.     Care has been taken to include sufficient number of modes so that the model reproduces many features found in experiments and simulations.  Also, we ensure that the model results are reasonably close to the simulation results.   The vertical velocity field ($v_{3}$) and vertical vorticity field ($\omega_{3}$) in terms of the chosen modes are
\begin{eqnarray}
v_3 &=& W_{101}(t)\cos(k x)\sin(\pi z) + W_{011}(t)\cos(k y)\sin(\pi z)\nonumber\\
& &+W_{202}(t)\cos(2 k x)\sin(2 \pi z) + W_{022}(t)\cos(2 k y)\sin(2 \pi z)\nonumber\\
& &+W_{103}(t)\cos(k x)\sin(3 \pi z) + W_{013}(t)\cos(k y)\sin(3 \pi z)\nonumber\\
& &+W_{301}(t)\cos(3 k x)\sin(\pi z) + W_{031}(t)\cos(3 k y)\sin(\pi z)\nonumber\\
& &+W_{121}(t)\cos(k x)\cos(2 k y)\sin(\pi z)+W_{211}(t)\cos(2 k x)\cos(k y)\sin(\pi z) \nonumber\\
& &+W_{112}(t)\cos(k x)\cos(k y)\sin(2\pi z)+W_{111}(t)\sin(k x)\sin(k y)\sin(\pi z)\\
\omega_3 &=& Z_{100}(t)\cos(k x) + Z_{010}\cos(k y)\nonumber\\
& & + Z_{110}(t)\sin(k x)\sin(k y) + Z_{112}(t)  \sin(k x)\sin(k y) \cos(2 \pi z) \nonumber\\
& & + Z_{310}(t)\sin(3 k x)\sin(k y) + Z_{130}(t)\sin(k x)\sin(3 k y)\nonumber\\
& & + Z_{120}(t)\cos(k x)\cos(2 k y) + Z_{210}(t)\cos(2 k x)\cos(k y)\nonumber\\
& & + Z_{102}(t)\cos(k x)\cos(2 \pi z) + Z_{012}(t)\cos(k y)\cos(2 \pi z)\nonumber\\
& & + Z_{201}(t)\cos(2 k x)\cos(\pi z) + Z_{021}(t)\cos(2 k y)\cos(\pi z)\nonumber\\
& & + Z_{111}(t)\cos(k x)\cos(k y)\sin(\pi z) + Z_{121}(t)\sin(k x)\sin(2 k y)\cos(\pi z)\nonumber\\
& &+Z_{211}(t)\sin(2 k x)\sin(k y)\cos(\pi z)
\end{eqnarray}
where $W_{lmn}$ and $Z_{lmn}$ are the Fourier amplitudes of the vertical velocity and vertical vorticity modes respectively with the three subscripts ($l, m, n$) indicating the wavenumber components along the $x$, $y$, and $z$ directions respectively.   The modes $(1,0,1)$ and $(0,1,1)$ are the most important modes of our model, and they represent the rolls along $y$ and $x$ directions respectively.   For the square pattern, the most important participating triad is $\{ (1,0,1), (0,1,1), (1,1,2) \}$.  Note that the wavenumbers of the interacting triad satisfy ${\bf k}={\bf p}+{\bf q}$.

The horizontal components of the velocity field can be computed using the incompressibility condition of the velocity field (Eq.~(\ref{eq:continuity})), and the temperature field $\theta$ can be computed using Eq.~(\ref{eq:theta}).  A Galerkin projection of the RBC equations~(\ref{eq:v3}-\ref{eq:omega3}) on the above modes provides a  set of $27$-dimensional coupled first-order nonlinear ordinary differential equations for the amplitudes of the above Fourier modes. On this 27-mode model, we perform a detailed bifurcation analysis that we describe in the subsequent sections. 

Our 27-mode model is a superset of the 13-mode model of \cite{Pal:EPL_2009}.  The nonlinear interactions of the 13-mode model are indicated by blue curves in Fig.~\ref{fig:modes_lowdim}.  Additional interactions induced by the new modes of the 27-mode model are represented by red curves in the figure.   A primary motivation for the 27-mode model is to be able to generate wavy rolls.  The triads $ \{ (0,1,0), (1,0,1), (1,1,1) \}$  and $ \{ (1,0,0), (0,1,1), (1,1,1) \}$  play a critical role in  inducing wavy rolls along the $y$ and $x$ axes respectively.    In the present paper we will investigate the dynamics of these wavy rolls in RBC using  numerical simulations and the 27-mode model  along with other features that are generated by the inclusion of these modes.

\begin{figure}
\begin{center}
\includegraphics[height=!,width=12cm]{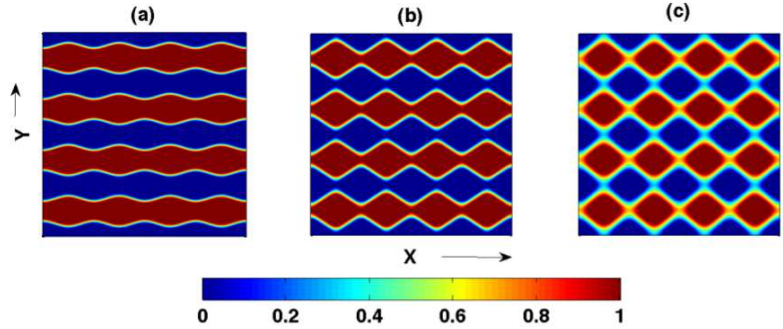}
\end{center}
\caption{Oscillating asymmetric square (OASQ) pattern in the mid-plane of the convection box in zero-P RBC. The pattern is obtained from the 27-mode model at $r=1.1$. Snapshots at: (a) t = 0, (b) t = T/4, and (c) t = T/2, where T is the time period of oscillation.} \label{fig:pattern_OASQ}
\end{figure}

\begin{figure}
\begin{center}
\includegraphics[height=!,width=12cm]{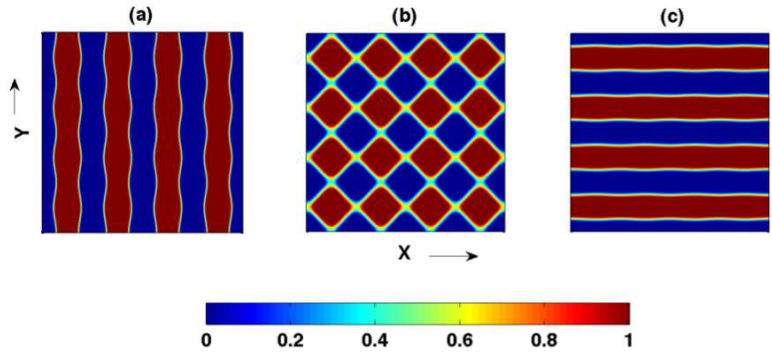}
\end{center}
\caption{Relaxation oscillations with an intermediate square regime (SQOR) pattern observed in the 27-mode model at $r=1.05$. Snapshots at: (a) t = 0, (b) t = T/4, and (c) t = T/2.} \label{fig:pattern_SQOR}
\end{figure}

\begin{figure}
\begin{center}
\includegraphics[height=!,width=12cm]{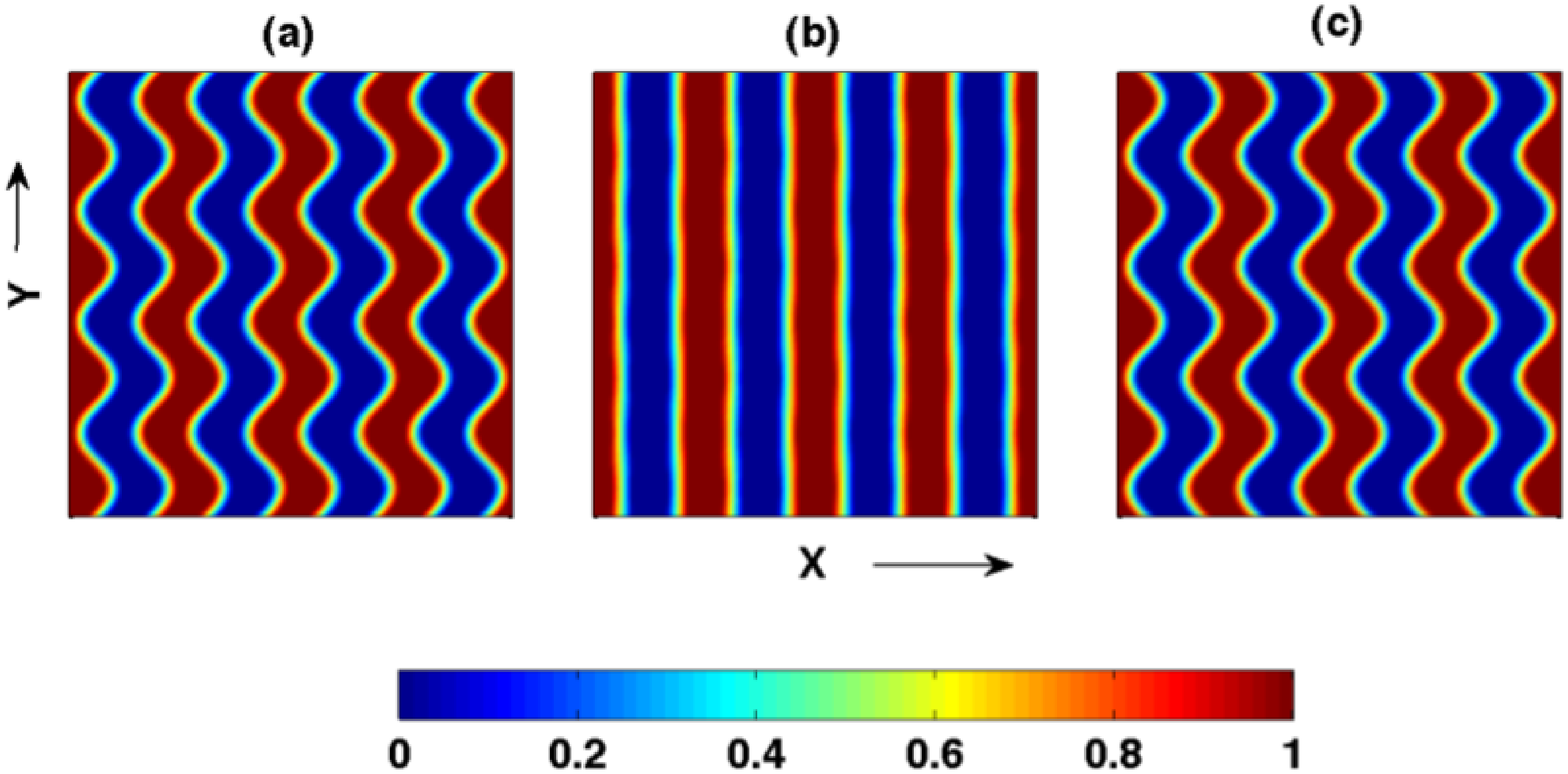}
\end{center}
\caption{Wavy roll pattern in the 27-mode model at $r=1.15$. \break Snapshots at: (a) t = 0, (b) t = T/4, and (c) t = T/2.} \label{fig:pattern_wavyroll}
\end{figure}

We numerically solve the 27-mode model by employing accurate ODE solvers of MATLAB.  As a result we observe a variety of convective patterns: squares (SQ), asymmetric squares (ASQ), oscillating asymmetric squares (OASQ), relaxation oscillations with an intermediate square regime (SQOR), wavy rolls, chaotic squares, etc. We have illustrated three snapshots each of  OASQ in Fig.~\ref{fig:pattern_OASQ}, SQOR in Fig.~\ref{fig:pattern_SQOR},  and wavy rolls in Fig.~\ref{fig:pattern_wavyroll}.  For dynamics of these patterns as well as other patterns mentioned in this paper refer to the accompanying videos. Note that all the above patterns were also found in our DNS.  Earlier  \cite{thual:zeroP_1992} in his DNS of zero-P convection had shown the existence of SQ,  SQOR, oscillatory quasihexagons (SQOS), chaotic squares (SQCH), and chaotic quasihexagon (HXCH).   Thual observed the oscillatory and chaotic quasi-hexagons for Rayleigh numbers beyond the range investigated in this paper.

We investigate the origin of various convective flow patterns from the bifurcation diagrams generated using the low dimensional model. To generate the bifurcation diagram, we start first with the fixed points of the system. We compute the fixed points using the Newton-Raphson method for a given $r$, and these fixed points are subsequently continued using a fixed arc-length based continuation scheme for the neighbouring $r$ values~\cite[]{nanda,wahi}. The stability of the fixed points are ascertained through an eigenvalue analysis of the Jacobian.  New branches of fixed points and limit cycles are born when the eigenvalue(s) become zero (pitchfork) and purely imaginary (Hopf) respectively.  This process is continued on the new branch.   For aperiodic and chaotic solutions, we resort to numerical integration and report the extremum values of the important modes. We use our own MATLAB code as well as MATCONT~\cite[]{dhooge:matcont} for the analysis.

%%%%%%%%%%%%%%%%%%%%%%%%%%%%%%%%%%%
\section{Bifurcation analysis using model and simulation results }{\label{sec:model_results}}

In the present section we numerically solve Eqs.~(\ref{eq:v3}-\ref{eq:continuity}) using DNS and the 27-mode model in the range $0.98 \leq r \leq 1.25$.  This range of $r$ values is  near the onset of convection. We will  present the bifurcation diagrams associated with the different attractors using the low-dimensional model followed by a detailed comparison of the model results with those obtained from DNS. 

\begin{figure}
\begin{center}
\includegraphics[height=!,width=15cm]{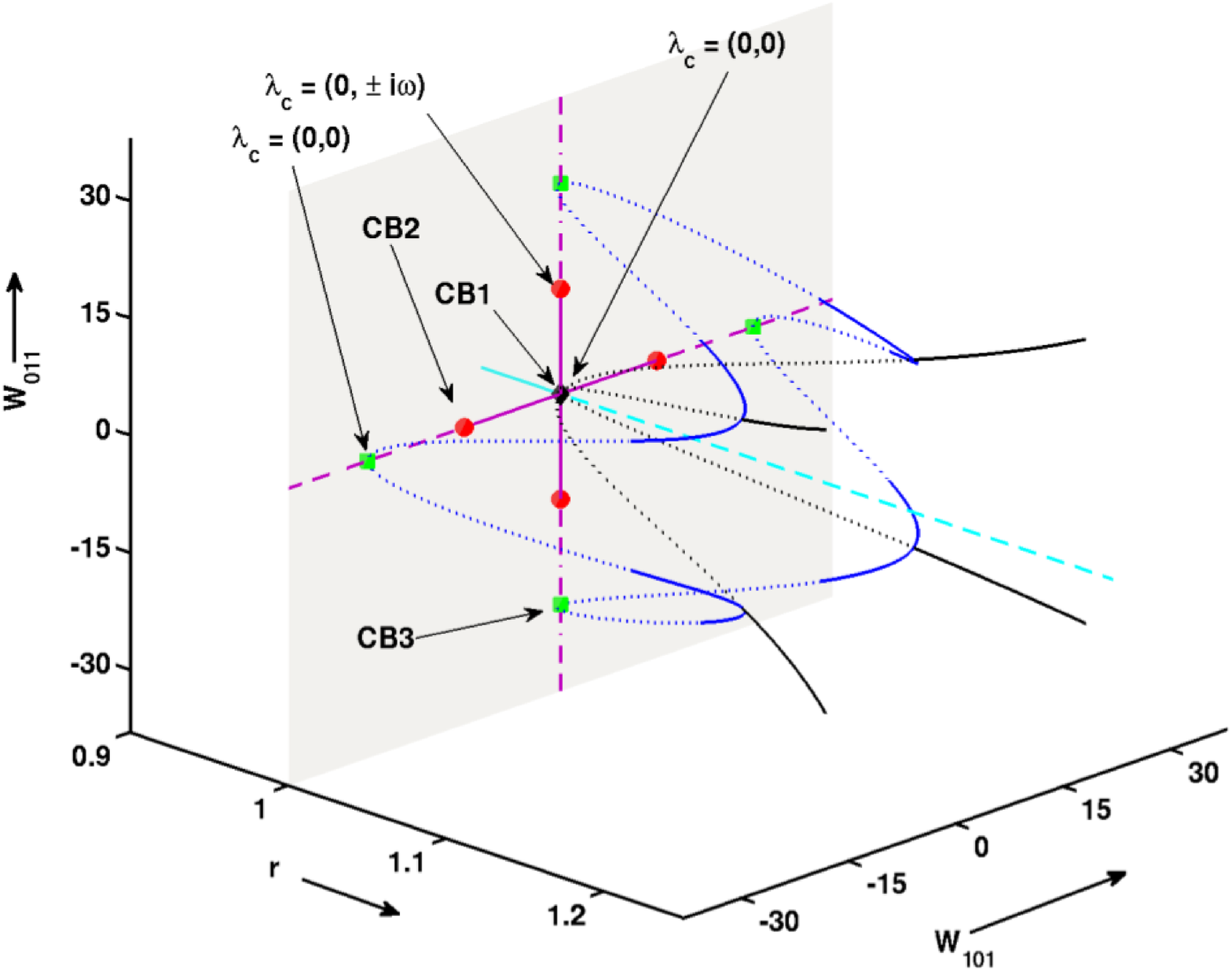}
\end{center}
\caption{Three dimensional view of the bifurcation diagram exhibiting the fixed points only. 
Solid and dashed curves represent the stable and unstable fixed points respectively. Black,
blue, and cyan curves represent stationary squares (SQ), asymmetric stationary squares
(ASQ), and conduction state respectively. Solid purple lines on the $W_{101}$ and $W_{011}$ axes represent the neutrally stable 2D roll solutions and chained purple lines represent unstable 2D rolls. CB1 (black diamond), CB2 (red dots) and CB3 (green squares) are the codimension-2 bifurcation points on the $r=1$ plane.  The critical eigenvalues at CB1, CB2 and CB3 are denoted by $\lambda_{c}$ in the figure.}
\label{3d_bif}
\end{figure}

\subsection{Fixed points of the system}{\label{sec:fixed_points}}
Fixed points of a dynamical system and their bifurcations provide important clues about the
system dynamics. Therefore, we start our analysis by locating all
the fixed points of the 27-mode model.  In Fig.~\ref{3d_bif} we
display the projection of these fixed points on the
$W_{101}$-$W_{011}$ plane as a function of $r$.  For $r<1$, the only stable fixed point of the system is the origin, which
corresponds to the pure conduction state (cyan curve).  At $r=1$,
the conduction state loses stability, and neutrally stable pure roll
solutions (purple curves of Fig.~\ref{3d_bif}) and four unstable branches of symmetric square solutions (dotted black curves of Fig.~\ref{3d_bif}) satisfying $|W_{101}| = |W_{011}|$ are born.  Note
that the stability matrix at the origin (CB1) has double zero eigenvalues,
and hence CB1 is a codimension-2 bifurcation point. 

The neutrally stable 2D roll solutions of the $r=1$ plane become unstable
through another codimension-2 bifurcation  at CB2 \{($W_{101} \simeq \pm 13.44$, $W_{011}=0$); ($W_{011} \simeq \pm 13.44$, $W_{101}=0$)\}, which is represented by large red dots on the $r=1$ plane.  The stability matrix at CB2 has a zero eigenvalue and a purely imaginary pair ($\lambda_c = (0, \pm i \omega$) with $\omega \approx 14.2$).     As a consequence of the complex eigenvalues of CB2,  for $r>1$ periodic solutions are born, and the 2D rolls lose their stability (chained purple lines of Fig.~\ref{3d_bif}). These
periodic solutions are also unstable akin to the  symmetric square fixed points associated with CB1.  We will show later that the wavy rolls are associated with CB2.  Note that CB2 and their associated attractors are absent  in the 13-mode model~\cite[]{Pal:EPL_2009}. 

The unstable roll solutions which persist at $r=1$ with amplitudes greater than $13.44$ subsequently undergo yet another codimension-2
bifurcation at CB3  \{($W_{101}=\pm 26.94, W_{011}=0$); ($W_{101}=0, W_{011}=\pm 26.94$)\}, which are represented as large green
squares on the $r=1$ plane.  The stability matrix at CB3 has double zero eigenvalues ($\lambda_c=0,0$).   Consequently unstable asymmetric square solutions with $|W_{101}| \neq |W_{011}|$ (dotted blue curves) are born.  This bifurcation is also present in the 13-mode model~\cite[]{Pal:EPL_2009}, and the attractors associated with this bifurcation in the 13-mode model carry over to the 27-mode model as well.   Note that the ASQ solutions of the 27-mode model have two pairs of unstable eigenvalues  as opposed to one unstable pair in the 13-mode model due to the presence of CB2.   

For low-P convection \cite{mishra:EPl_2010} have earlier observed that the 2D rolls undergo a pitchfork bifurcation followed by a Hopf.  In the limiting case of zero-P, the Hopf bifurcation point merges with the pitchfork bifurcation point  at $r=1$,  and the critical eigenvalues at this bifurcation point are $(0,0)$ giving rise to a codimension-2 bifurcation CB3. With an increase in $r$, the double zero eigenvalues split into unstable complex conjugate pair giving rise to a scenario very similar to the Takens-Bogdanov bifurcation \cite[]{guckenheimer:book,kuznetsov:book}.

As shown in Fig.~\ref{3d_bif}, on the projection of $W_{101}$-$W_{011}$ plane, there are $13$ unstable fixed points for $r>1$:  one corresponding to the pure conduction state, four satisfying $|W_{101}| = |W_{011}|$ (SQ),  and the remaining eight satisfying $|W_{101}| \ne |W_{011}|$ (ASQ).   After two successive inverse Hopf bifurcations, to be described later, the unstable ASQ fixed points become stable.  These
stable fixed points are shown by the solid blue curves in
Fig.~\ref{3d_bif}. Subsequently at $r\simeq 1.1690$, these stable
ASQs merge (via a pitchfork bifurcation) with the symmetric square
solutions that originate from CB1 and stabilize them (solid black
curves in Fig.~\ref{3d_bif}). Note that there is a small difference in the values
of $r$ corresponding to the stabilization of the ASQ solutions and
SQ solutions  for the 13-mode model  and the 27-mode model. 
In summary, the 3D figure (Fig.~\ref{3d_bif}) is
qualitatively similar to the corresponding figure of the 13 mode
model~\cite[]{Pal:EPL_2009} except the codimension-2 bifurcation  at CB2 which is responsible for the wavy rolls.  In the following discussions we will describe the bifurcation diagrams including limit cycles, chaotic attractors etc.

Bifurcation diagrams of the 27-mode model are quite complex.  They include six different types of chaotic attractors, various types of fixed points and periodic solutions.  The model also has multiple coexisting attractors for a given value of $r$.   To disentangle its complexity, we present the bifurcation diagrams as four separate diagrams, ``Bif-13M'', ``Bif-A'', ``Bif-B'', ``Bif-C'', that highlight different features of the dynamics.   First we draw the bifurcation diagram ``Bif-13M'' associated with the 13-mode model~\cite[]{Pal:EPL_2009},  which is a subset of the 27-mode model.  Later, we will contrast  the bifurcation diagram of the  27-mode model (denoted by ``Bif-A'') with  the diagram of the 13-mode model.

\begin{figure}
\begin{center}
\includegraphics[height=!,width=12cm]{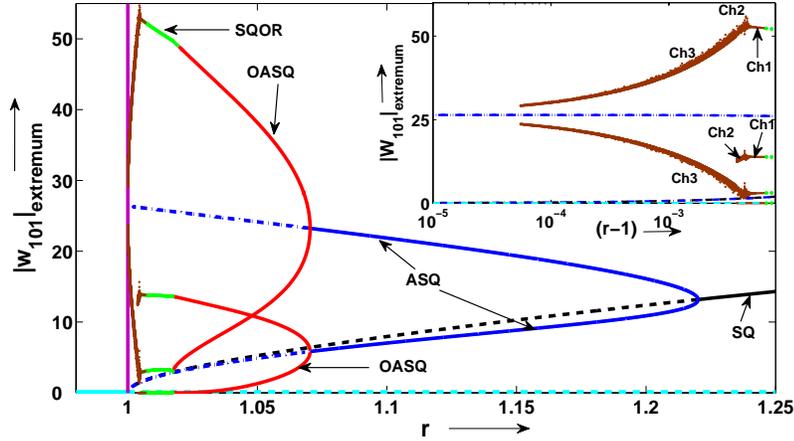}
\end{center}
\caption{Bifurcation diagram Bif-13M of the 13-mode model for $0.98 \leq r
\leq 1.25$~\cite[adapted from][]{Pal:EPL_2009}. The stable branches corresponding to SQ and ASQ are represented by  solid black and blue lines respectively. Red, green, and brown points represent the extrema of OASQ, SQOR, and chaotic solutions respectively.  A zoomed view of the bifurcation diagram for the chaotic regime is shown in the inset. Branches corresponding to the unstable fixed points are represented by dashed lines. Cyan line represents the conduction state.}
\label{fig:bifur13}
\end{figure}
%%%%%%%%%%%%%%%%%%%%%%%%%%%%%%

\subsection{Bifurcation diagram  of the 13-mode model}{\label{sec:bif13M}}

As illustrated by Fig.~\ref{fig:modes_lowdim}, the 13-mode model is a subset of the 27-mode model.  If we force only the modes of the 13-mode model to be nonzero, and others to be zero, naturally the bifurcation diagram corresponding to the 13-mode model is reproduced.  \cite{Pal:EPL_2009} contains a detailed discussion on this diagram and the associated flow patterns (both from DNS and the model).  Here we only provide a brief description. 

The bifurcation diagram for the 13-mode model is illustrated in Fig.~\ref{fig:bifur13} in which we plot the positive value of $|W_{101}|_{extremum}$ as a function of $r$.   For zero-P convection, chaos is observed at the onset itself.  Therefore, ~\cite{Pal:EPL_2009} start  their analysis at $r=1.4$ where symmetric square (SQ) patterns are observed.  These states are represented by the solid black curve of Fig.~\ref{fig:bifur13} (here the diagram is shown only for $r\le 1.25$).  At around $r \approx 1.2201$, SQ branches bifurcate to ASQ solutions through a supercritical pitchfork bifurcation. The solid blue curves of Fig.~\ref{fig:bifur13} correspond to ASQ patterns.  The ASQ branches bifurcate to OASQ solutions (the solid red curves) through a Hopf bifurcation.   The limit cycles thus generated grow in size and touch the saddles (dashed line of the SQ branch) to create a very narrow window of homoclinic chaos.  After this, the system again becomes periodic (SQOR) with the merger of the limit cycles.  The SQOR patterns transform to a chaotic attractor {\em Ch1} through a homoclinic bifurcation.  {\em Ch1} turns to {\em Ch2} and subsequently to {\em Ch3} through ``crisis''.  \cite{Pal:EPL_2009}  observed these patterns in both model and DNS.  Earlier, \cite{thual:zeroP_1992} had observed SQ, ASQ, and SQOR patterns in his DNS runs.

In the subsequent subsections we will describe the bifurcation scenario for the 27-mode model.

%%%%%%%%%%%%%%%%%%%%%%%%%%%%%%
\subsection{Bifurcation diagram Bif-A of the 27-mode model}{\label{sec:Bif-A}}

\begin{figure}
\begin{center}
\includegraphics[height=!,width=12cm]{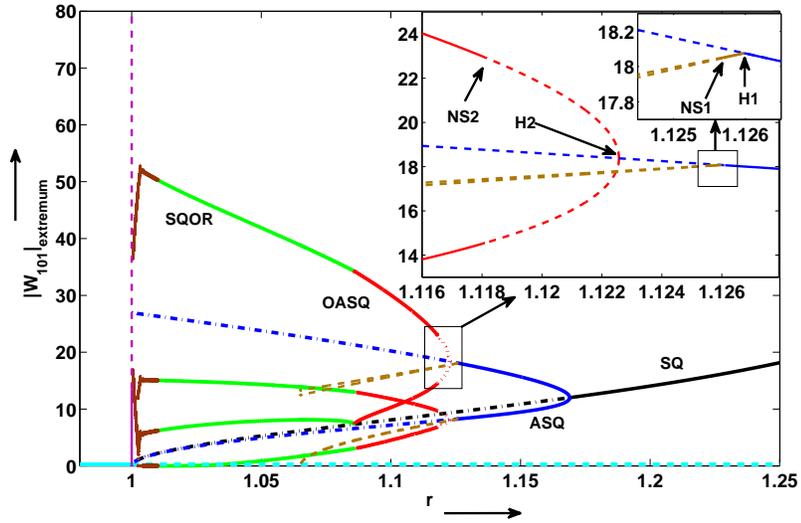}
\end{center}
\caption{Bifurcation diagram Bif-A of the 27-mode model with the same color convention as Bif-13M shown in Fig.~\ref{fig:bifur13}.  This diagram is qualitatively similar to Bif-13M.  New features of Bif-A are the H1, NS1, H2, and NS2 bifurcations shown in the boxed region ($1.116 < r < 1.128$) whose zoomed view is shown in the inset.} 
\label{fig:bifur27A}
\end{figure}

The square pattern described above is also observed in the 27-mode model.  However, for SQ in the 27-mode model, twelve modes $W_{111}$, $Z_{110}$, $Z_{112}$, $Z_{100}$, $Z_{010}$, $Z_{111}$, $Z_{210}$, $Z_{120}$, $Z_{201}$, $Z_{021}$, $Z_{102}$, and $Z_{012}$ still remain zero.  When we continue the SQ branch of the 27-mode model, we obtain a new bifurcation-diagram called ``Bif-A"  shown in Fig.~\ref{fig:bifur27A}.   The
bifurcation diagram Bif-A is qualitatively similar to Bif-13M except in a  narrow window of $ 1.116 \le r \le 1.128$ where additional bifurcations are observed. The ASQ branch of solutions in Bif-A have seventeen active modes with the modes  $W_{111}$, $Z_{100}$, $Z_{010}$, $Z_{111}$, $Z_{210}$, $Z_{120}$, $Z_{201}$, $Z_{021}$, $Z_{102}$, and $Z_{012}$ as zeros.

\begin{figure}
\begin{center}
\includegraphics[height=!,width=12cm]{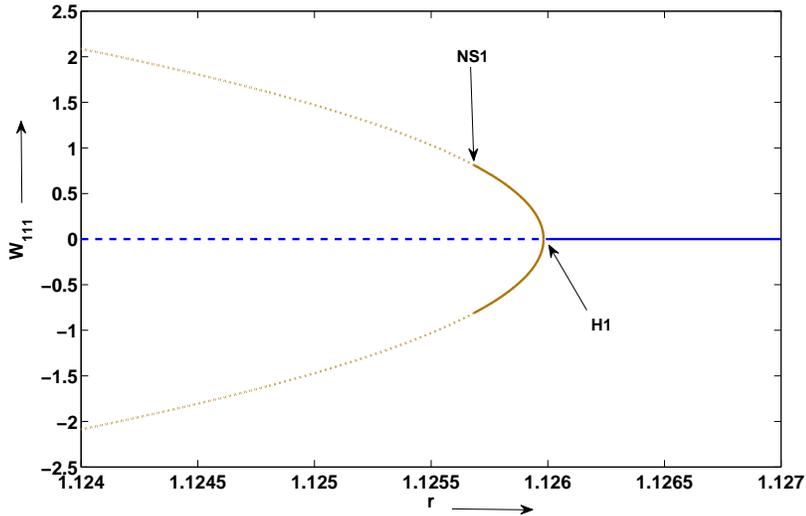}
\end{center}
\caption{Plot of $W_{111}$ vs. $r$ near the first Hopf bifurcation (H1) of the
ASQ branch. The solid brown curve represents the limit cycles generated after the first Hopf (H1).}
\label{fig:r_w111_projection}
\end{figure}

\begin{figure}
\begin{center}
\includegraphics[height=!,width=12cm]{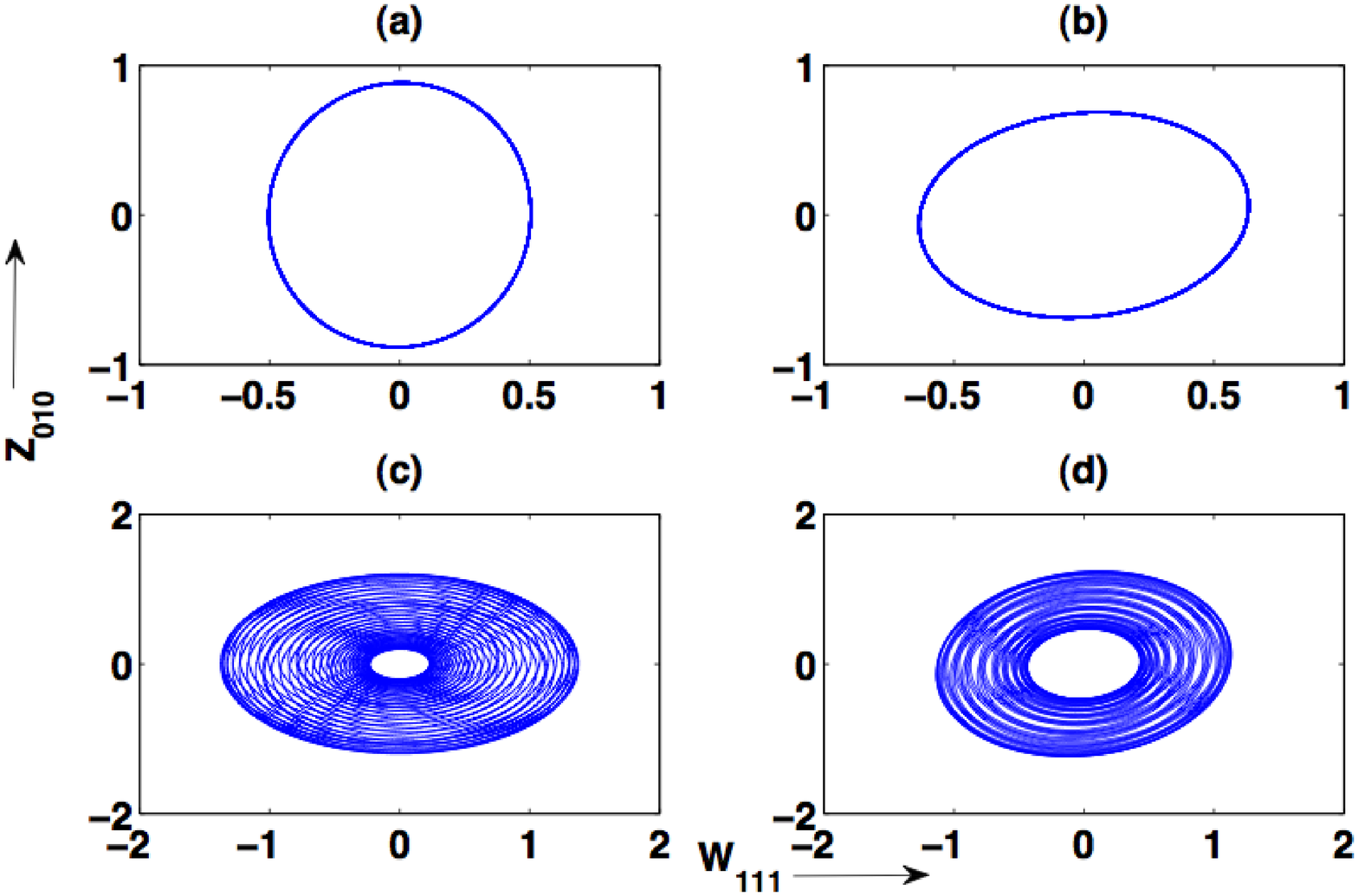}
\end{center}
\caption{Projection of the phase space on the $W_{111}$ - $Z_{010}$
plane. The limit cycles born through H1 in (a) DNS at r=1.138 and (b)
model at r=1.1258. Quasiperiodic attractor born through NS1 in (c)
DNS at $r=1.131$ and (d) model at $r=1.1245$.} 
\label{fig:H1_NS1}
\end{figure}

The new features of Bif-A are as follows.  At  $r = 1.1260$, the ASQ branch undergoes a  supercritical Hopf bifurcation (H1, see Fig.~\ref{fig:r_w111_projection}) resulting in a time-periodic convective flow as illustrated in Fig.~\ref{fig:H1_NS1}(a,b),  where we show a projection of the limit cycle obtained from the DNS and the model on the $W_{111}$-$Z_{010}$ plane.   All the 27 modes are active for these periodic flow patterns.

\begin{figure}
\begin{center}
\includegraphics[height=!,width=12cm]{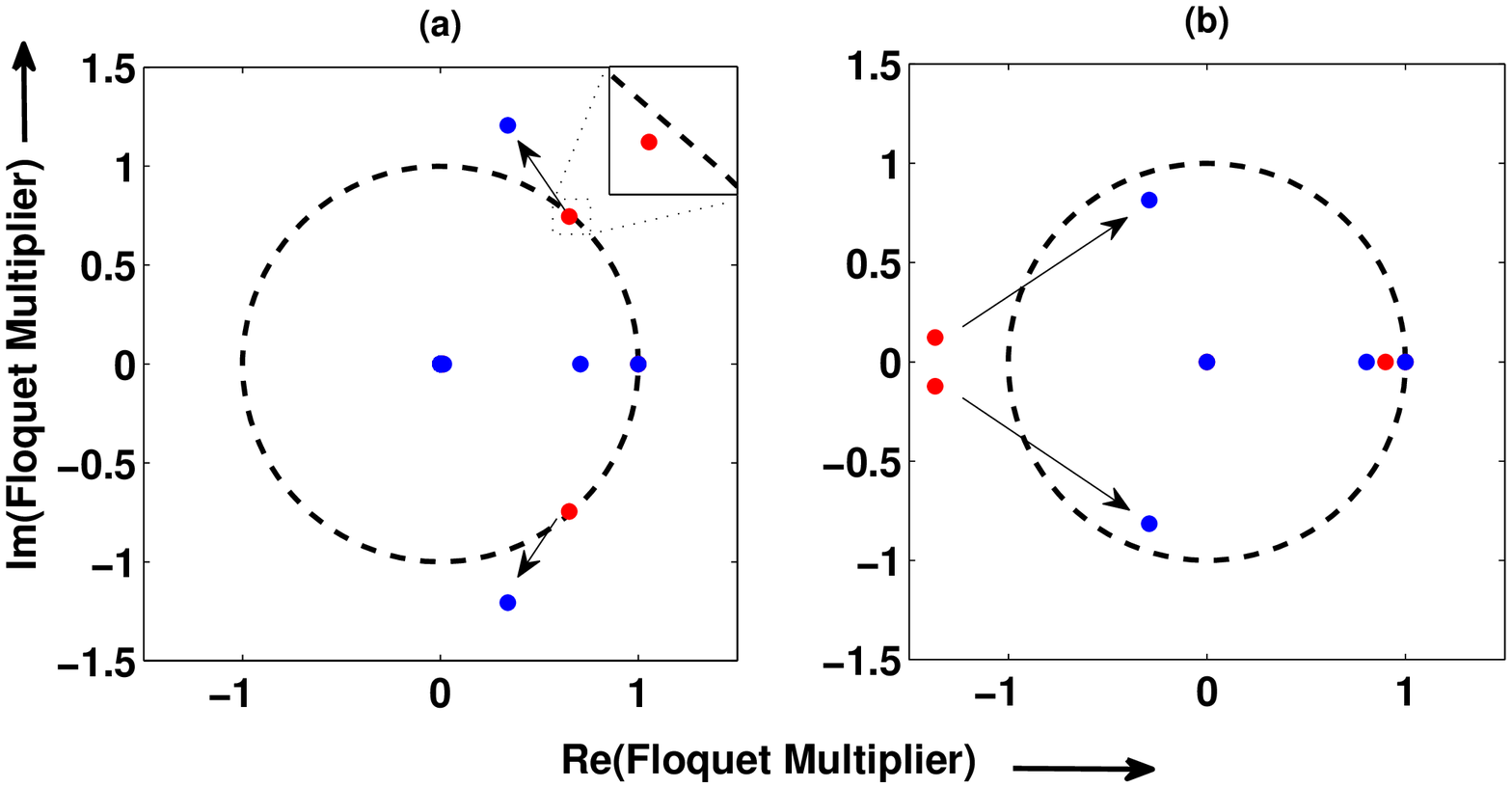}
\end{center}
\caption{Movement of Floquet multipliers in the complex plane during NS1 and NS2 bifurcations. (a) At NS1, a pair of complex Floquet multipliers move out of the unit circle (red dots changing to blue). (b) At NS2, a pair of complex Floquet multipliers  enter the unit circle (red dots again becoming blue).} 
\label{fig:Floquet_NS1_NS2}
\end{figure}

As $r$ is reduced further, at $r=1.1257$ a new frequency incommensurate with the original frequency is born through a Neimark-Sacker bifurcation (NS1) and the limit cycle becomes unstable.  Here, a pair of imaginary Floquet multipliers cross the unit
circle outwards as illustrated in Fig.~\ref{fig:Floquet_NS1_NS2}(a).   The phase space
trajectory of the system on the $W_{111}$-$Z_{010}$ plane is therefore
quasiperiodic as demonstrated in Fig.~\ref{fig:H1_NS1}(c,d) for
the DNS and the model respectively. The unstable limit cycle 
continues till $r=1.0651$ where they meet the unstable limit cycles
of the wavy rolls, which will be discussed in~\S~\ref{sec:Bif-C}.

\begin{figure}
\begin{center}
\includegraphics[height=!,width=12cm]{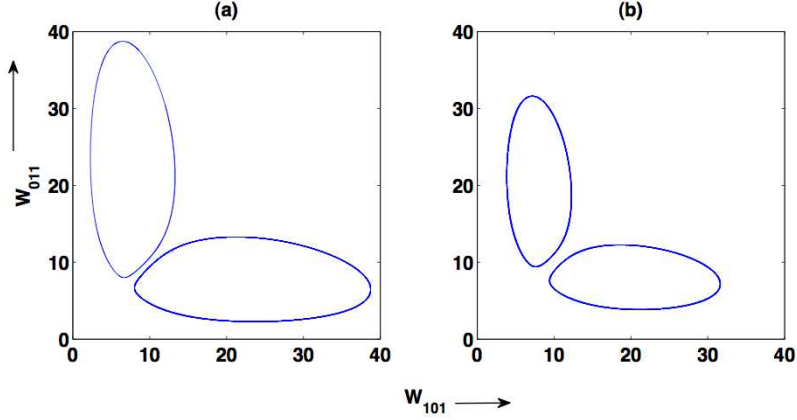}
\end{center}
\caption{Projections of the two coexisting OASQ solutions on the $W_{101}$ -$W_{011}$ plane just after the NS2 bifurcation point: (a) in DNS at $r=1.0768$ and (b) in the model at $r=1.095$.}
\label{fig:NS2}
\end{figure}

On further reduction of $r$, at $r= 1.1226$, another Hopf bifurcation
(H2) takes place on the unstable ASQ branch. The limit cycle born
from H2 is however unstable. At $r = 1.1181$, this limit cycle becomes stable via
an inverse Neimark-Sacker bifurcation (NS2) wherein the unstable
Floquet multiplier pair enters the unit circle as evident in
Fig.~\ref{fig:Floquet_NS1_NS2}(b). The resulting stable limit cycle
is the oscillatory asymmetric square (OASQ) solution of the 13-mode
model. Figure \ref{fig:NS2} shows a projection of the limit cycle
corresponding to the OASQ solution obtained from the DNS and
the model on the $W_{101}$-$W_{011}$ plane.  The
quasiperiodic solutions exist only in the range
$r=1.1181$-$1.1257$, i.e., between NS1 and NS2 and disappears after NS2.  Note that the  attractors  between H1 and NS2  contain all the 27 modes, but  beyond NS2 the ten modes $W_{111}$, $Z_{100}$, $Z_{010}$, $Z_{111}$, $Z_{210}$, $Z_{120}$, $Z_{021}$, $Z_{102}$, and $Z_{012}$ again become zero.

Beyond NS2, the patterns and associated bifurcations from OASQ to SQOR
to the chaotic attractors {\em Ch1, Ch2} and {\em Ch3} in the decreasing $r$
direction are exactly the same as for Bif-13M.    The range of $r$  corresponding to these patterns are approximately $r=1.086$--$1.1181$  for OASQ, $r=1.0046$--$1.086$ for SQOR, $r=1.0034$--$1.0046$ for Ch1, $r=1.0025$--$1.0034$ for {\em Ch2} and $r=1$--$1.0025$ for {\em Ch3}.  Comparison with the 13-mode model indicates that the ranges of $r$ for the above patterns as well as that for SQ and ASQ are different.  This is due to the fact that more than 13 modes are active in the present model for Bif-A.  The 27-mode model reproduces the ranges of the patterns obtained from DNS more accurately.  Note that the modes absent  for SQ, ASQ, OASQ, SQOR, {\em Ch1} to {\em Ch3} flow patterns are active for wavy rolls that will be described in~\S\ref{sec:Bif-C}.

%%%%%%%%%%%%%%%%%%%%%%%%%%%%%%
\subsection{Bifurcation diagram Bif-B of the 27-mode model}{\label{sec:Bif-B}}

\begin{figure}
\begin{center}
\includegraphics[height=!,width=12cm]{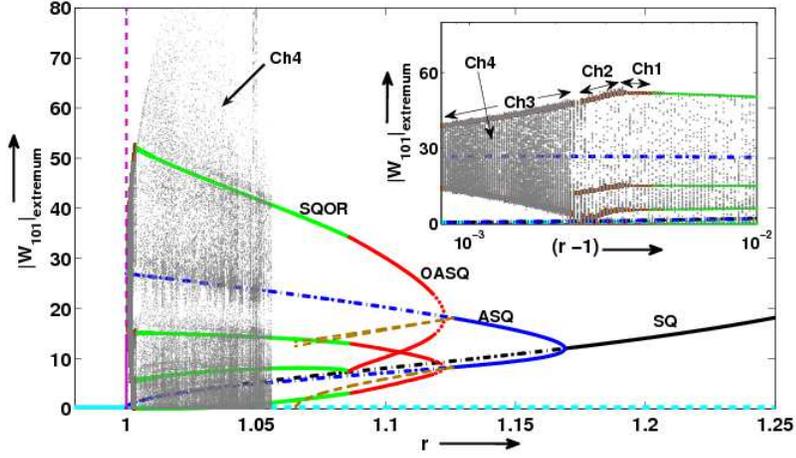}
\end{center}
\caption{Bifurcation diagram Bif-B with the same color convention as Bif-13M (Fig.~\ref{fig:bifur13}). A large chaotic attractor represented by gray dots {\em Ch4} is born at CB3. The chaotic attractor {\em Ch4} coexists with {\em Ch1}, {\em Ch2}, {\em Ch3} (shown in the inset using brown dots), and SQOR.}
\label{fig:bifur27B}
\end{figure}

\begin{figure}
\begin{center}
\includegraphics[height=!,width=12cm]{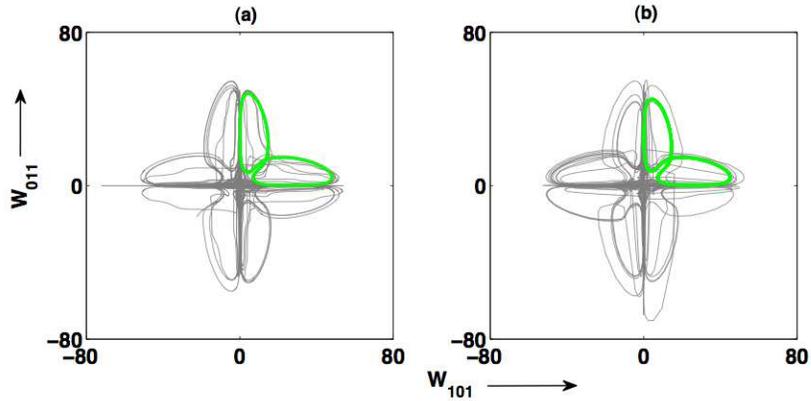}
\end{center}
\caption{Phase space projection of coexisting SQOR (green) and chaotic attractor {\em Ch4} (gray) on the $W_{101}$-$W_{011}$ plane at $r = 1.0342$: (a) obtained from model and (b) obtained from DNS. }
\label{fig:multi_attrac}
\end{figure}

When we start from an arbitrary initial condition for $r>1$ near the onset, most often the system tends to another attractor {\em Ch4}, which differs from {\em Ch1, Ch2}, and {\em Ch3}.    The codimension-2 bifurcation point CB3 of Fig.~\ref{3d_bif} generates the chaotic attractor {\em Ch4} as well.  The bifurcation diagram ``Bif-B" shown in Fig.~\ref{fig:bifur27B} contains the attractor {\em Ch4}.   The attractor {\em Ch4} coexists with the chaotic attractors {\em Ch1, Ch2}, {\em Ch3} and SQOR for $r=1$--$1.056$.  This feature is illustrated in Fig.~\ref{fig:multi_attrac} where a phase space projection (both from the model  and DNS) for two different initial conditions at $r = 1.0342$ yield the SQOR (green curve) and the {\em Ch4} (grey trajectory) attractors. Clearly the trajectories of {\em Ch4} explore all the four quadrants of the $W_{101}$-$W_{011}$ plane.   The qualitative behavior of the {\em Ch4} attractor is similar to {\em Ch2}, but its size is larger than {\em Ch2} \cite[compare with Fig. 5 of][]{Pal:EPL_2009}.

Note that there are two complex pairs of unstable eigenvalues associated with ASQ.  The first one carries from CB2, and the second one is due to the splitting of the double zero eigenvalues at CB3 into a complex pair (see~\S\ref{sec:fixed_points}).  The {\em Ch4} attractor is associated with the first pair, while the {\em Ch3} attractor is related to the second pair.    The disappearance of the chaotic attractor {\em Ch4} is possibly through a boundary crisis \cite[]{hilborn:book} wherein an unstable periodic
solution hits the basin boundaries of {\em Ch4}. This unstable periodic
solution possibly has connections with the unstable periodic
solutions originating from the branch point CB2 of Fig.~\ref{3d_bif}. 

In the next subsection, we will discuss the bifurcations 
associated with the solutions arising from the branch point CB2 whose flow patterns resemble the wavy rolls.

%%%%%%%%%%%%%%%%%%%%%%%%%%%%%

\subsection{Bifurcation diagram Bif-C of the 27-mode model}{\label{sec:Bif-C}}

\begin{figure}
\begin{center}
\includegraphics[height=!,width=12cm]{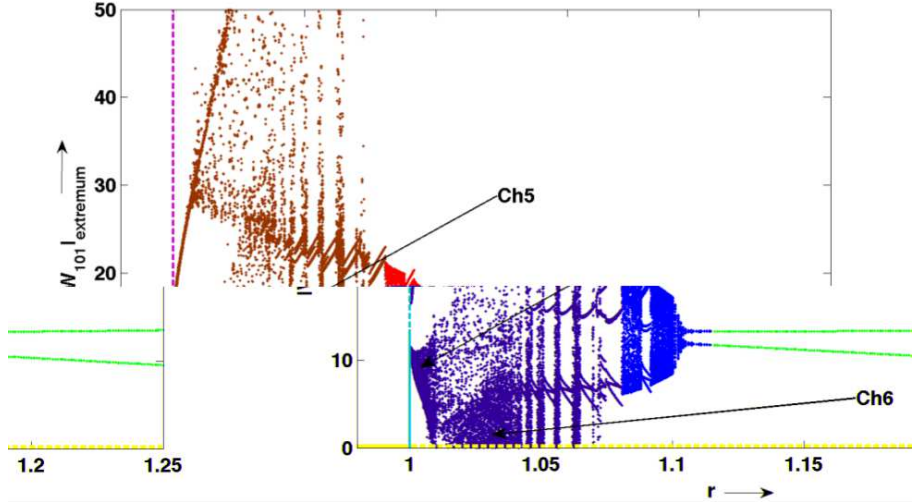}
\end{center}
\caption{Bifurcation diagram Bif-C of the wavy roll solutions. The blue, red, and brown points represent periodic, quasiperiodic and chaotic ({\em Ch5} and {\em Ch6}) solutions respectively. }
\label{fig:bifur3}
\end{figure}

\begin{figure}
\begin{center}
\includegraphics[height=!,width=12cm]{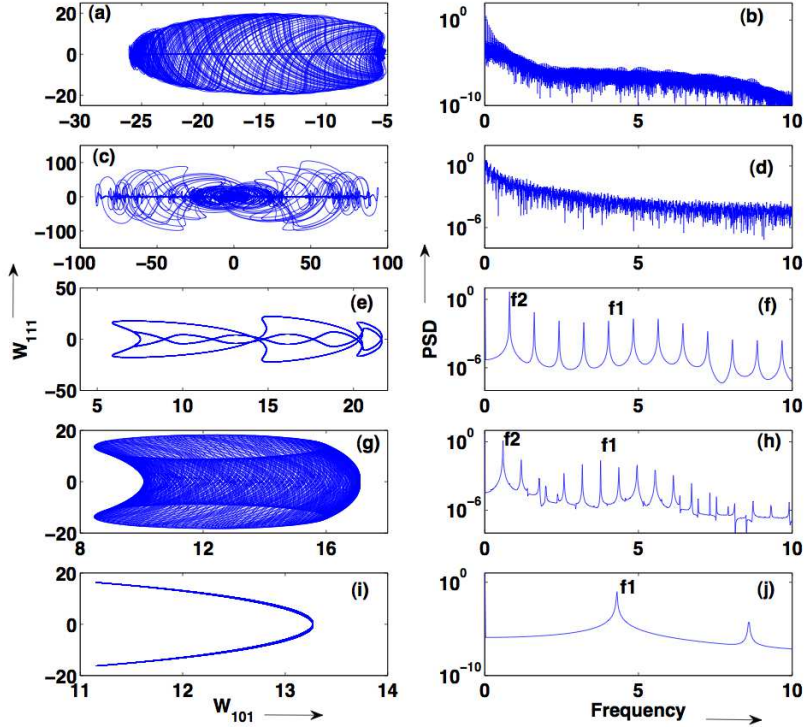}
\end{center}
\caption{Phase space projections on the $W_{101}$-$W_{111}$  plane for the wavy rolls and their corresponding power spectra obtained from the model:  (a,b) the chaotic attractor {\em Ch5}  at $r = 1.005$;  (c,d) the chaotic attractor {\em Ch6}  at $r = 1.05$;  (e,f) frequency locked state with $f1/f2 \approx 5$ at $r = 1.078$;  (g,h) quasiperiodic state with $f1/f2\approx 6.33$ at $r = 1.10$;  (i,j) periodic state at $r = 1.15$ ($f1=4.3$). }
\label{fig:roll_QP_model}
\end{figure}

The 27-mode model has another chaotic attractor near the onset that originates from the bifurcation point CB2 of Fig.~\ref{3d_bif}.  Recall from \S\ref{sec:fixed_points} that CB2 is a codimension-2 bifurcation point whose stability matrix has a simple zero eigenvalue and an imaginary pair $(0,\pm i \omega)$.  As a consequence, an unstable limit cycle is generated as $r$ is increased beyond 1. The attractors from this branch yield  periodic, quasiperiodic, and chaotic wavy rolls.   See Fig.~\ref{fig:pattern_wavyroll} for an illustration of a periodic wavy roll.   The diagram Bif-C (Fig.~\ref{fig:bifur3}) illustrates the bifurcation scenario of this type of solutions. The limit cycles generated through this bifurcation are unstable and they have an unstable torus associated with them as discussed in Section 7.4 of \cite{guckenheimer:book}. As a result, four chaotic attractors named {\em Ch5} are born.   A phase space projection of one of the {\em Ch5} attractor is shown in Fig.~\ref{fig:roll_QP_model}(a). Its chaotic nature is ascertained by the broad-band power spectrum exhibited  in Fig.~\ref{fig:roll_QP_model}(b).  As $r$ is increased further, the size of {\em Ch5} increases till $r \simeq 1.009$ after which a single large chaotic attractor {\em Ch6} is generated through an ``attractor-merging crisis''~\cite[]{hilborn:book}.   A phase space projection and power spectrum of {\em Ch6}  are shown in Fig.~\ref{fig:roll_QP_model}(c,d) respectively.  The chaotic attractors {\em Ch5} and {\em Ch6} are illustrated in the bifurcation diagram Bif-C.

\begin{figure}
\begin{center}
\includegraphics[height=!,width=8cm]{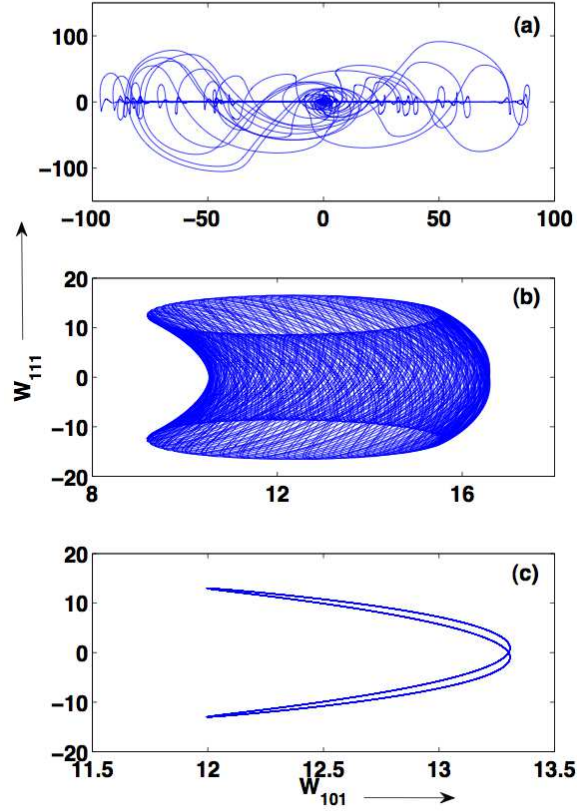}
\end{center}
\caption{Phase space projections on the $W_{101}$-$W_{111}$ plane
corresponding to the wavy rolls obtained from DNS: (a)
the chaotic attractor {\em Ch6} at $r=1.05$; (b) quasiperiodic state at $r=1.09$; (c) periodic state at $r=1.10$. } 
\label{fig:roll_QP_DNS}
\end{figure}

 As $r$ is increased further, we observe a series of phase-locked and stable quasiperiodic solutions.  Phase space projections and power spectra of the phase-locked ($r=1.078$) and quasiperiodic ($r=1.10$) states are shown in Fig.~\ref{fig:roll_QP_model}(e,f) and Fig.~\ref{fig:roll_QP_model}(g,h) respectively.  The phase-locked and quasiperiodic states are also evident in the bifurcation diagram Bif-C as banded and filled regions.  Further increase in $r$ transforms the quasiperiodic states to a limit cycle through an inverse Niemark-Sacker bifurcation.  A phase space projection and power spectrum of a limit cycle generated for $r=1.15$ are shown in Fig.~\ref{fig:roll_QP_model}(i,j).  Note that the transformations of phase-locked, quasiperiodic, and periodic states to one another can be understood by the movement of the Floquet multipliers of the underlying limit cycles (see \S \ref{sec:Bif-A}).  The above states have also been observed in DNS.  For example, Fig.~\ref{fig:roll_QP_DNS} illustrates chaotic, quasiperiodic, and periodic states obtained in DNS for $r=1.05, 1.09$, and $1.10$ respectively.
 
When we examine the active modes of Bif-C, we find that only the green colored modes of Fig.~\ref{fig:modes_lowdim} are active for these convective patterns.  The bifurcation diagram Bif-C has been generated by setting these modes as nonzero and all other modes as zero.  The most important among the active modes of Bif-C are $(1,0,1)$, $(0,1,0)$, and $(1,1,1)$ which are instrumental for the generation of wavy rolls along  the $y$ axis.  Naturally, the wavy rolls along the $x$ axis will have the complimentary set of modes, e.g., $(1,0,0)$ instead of $(0,1,0)$, etc.

Note that the above sets of solutions (Bif-A, Bif-B, Bif-C) are observed for different sets of initial conditions.  For both DNS and the model, a random initial condition generally produces Bif-B whose basin of attraction appears to be rather large.  If we keep only the green colored modes of Fig.~\ref{fig:modes_lowdim} as nonzero, we obtain Bif-C that correspond to the wavy rolls.  In fact, Bif-C is generated by first constructing a limit cycle for $r = 1.25$, and then continuing the solution for lower $r$ values.  The bifurcation diagram Bif-A is generated by starting from the SQ pattern for $r = 1.25$ for which all the modes except $W_{111}$, $Z_{110}$, $Z_{112}$, $Z_{100}$, $Z_{010}$, $Z_{111}$, $Z_{210}$, $Z_{120}$, $Z_{201}$, $Z_{021}$, $Z_{102}$, and $Z_{012}$ are nonzero.  Continuation of the above solution however generates ASQ solutions followed by H1, NS1, H2, and NS2 bifurcations where all the 27 modes are present.  

Wavy rolls are one of most studied convective patterns in experiments and numerical simulations.  The bifurcation diagram Bif-C provides a clear explanation for the origin of this pattern.  In the next section we will provide a quantitative comparison of the bifurcation results related to the wavy rolls with those from the experiments and previous simulations.

\section{Wavy rolls: a quantitative study}\label{sec:wavy_rolls}

 In this section we will analyze time scales of the wavy rolls of Bif-C quantitatively, and compare these values with some of the experimental and numerical results.   At the bifurcation point CB2, the stability matrix has a pair of complex eigenvalues $(0,\pm i\omega)$ with $\omega \approx 14.2$.  As a result, the unstable limit cycle originating from CB2 has a time period around $2 \pi/\omega \approx 0.44$ in units of $d^2/\nu$ (viscous time scale).  Subsequent periodic, quasiperiodic, and chaotic time series have time scales comparable to the above value since their origin is closely connected to the bifurcation CB2 at $r=1$ in Bif-C.   Our preliminary calculations indicate that the time period of oscillations for these patterns are within a factor of 10 from this value for $r=1$--$1.25$.
  
 Earlier \cite{krishnamurti:JFM_1973} observed time-dependent wavy rolls in her convection experiments on mercury ($P\approx 0.02$).  She observed multiple peaks with the time period ranging from 0.1 to 1 in the time units of $d^2/\nu$ \cite[see Fig.~3 of][]{krishnamurti:JFM_1973}.  Krishnamurti's experimental value for the time-scale of the chaotic wavy rolls is in the same range as our theoretical time-scale estimated above. \cite{willis:jfm_1970} and \cite{croquette:wavyroll_1989}  reported the time period of the oscillatory rolls using their experimental data for air ($P=0.7$) to be around 1 in the units of  $d^2/\nu$.   Using numerical simulations, \cite{lipps:jfm_1976} observed time periods of oscillatory rolls to be around $0.24$--$0.27$ in the units of $d^2/\kappa$ for $P=0.7$. \cite{meneguzzi:jfm_1987} found the period of the wavy oscillations to be around 0.065 viscous time units for $P=0.025$.  These results are in  general agreement (within a factor of 10) with our theoretical finding based on the bifurcation analysis.   Note that \cite{busse:JFM_1972} reported  that the time period  of the oscillatory instability in the units of $d^2/\nu$ is independent of the Prandtl number.  Hence the time-scales are not expected to vary appreciably even when we change the Prandtl number by  an order of magnitude which is consistent with the results for mercury ($P\approx 0.02$) and air ($P\approx 0.7$). Therefore, a comparison of our results for $P=0$ with those obtained for finite $P$ is also justified. 
 
\begin{figure}
\begin{center}
\includegraphics[height=!,width=12cm]{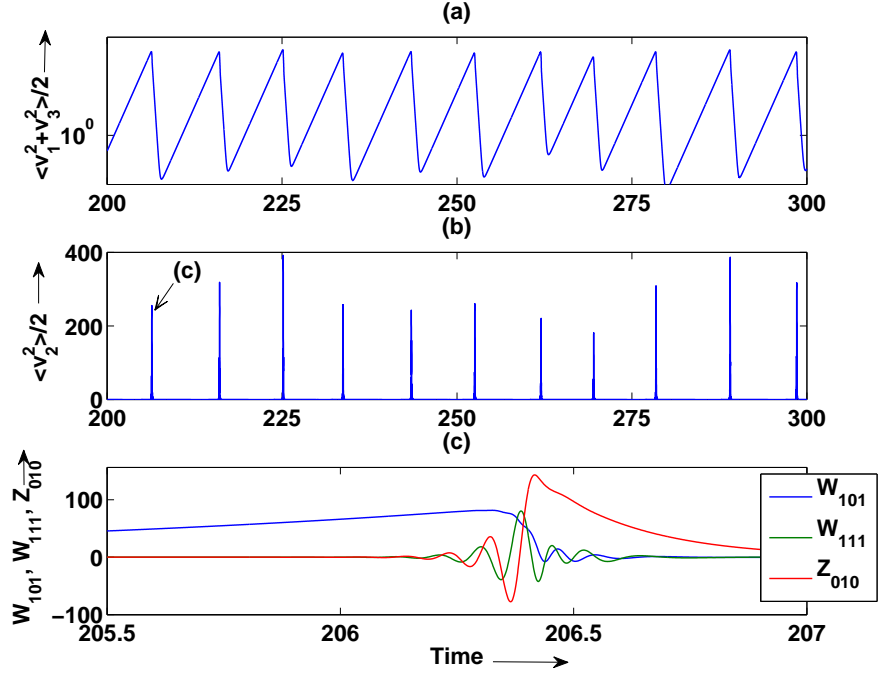}
\end{center}
\caption{Time series of $\langle v_1^2 + v_3^2 \rangle / 2$ (panel (a)) and $\langle v_2^2 \rangle$/2 (panel (b)) obtained from the model at $r=1.05$. Panel (c) shows the time series of the modes $W_{101}$, $W_{111}$, and $Z_{010}$ during critical bursting.} \label{fig:wavy_roll_energy}
\end{figure}

Oscillatory instabilities and their saturation through critical bursting have been studied by Kumar and coworkers \cite[]{kumar:JP_1996,kumar:burst_2006} using several low-dimensional models.  They show that the growth of the mode $W_{101}$ is saturated by the vorticity mode $Z_{010}$.   In Fig.~\ref{fig:wavy_roll_energy}(a,b) we plot the time series of $\langle v_1^2 + v_3^2 \rangle$ (sum of kinetic energy along $x$ and $z$ axes) and $\langle v_2^2 \rangle$ (kinetic energy along $y$ axis) computed from our 27-mode model for $r=1.05$.  These results are in general agreement with the results of Kumar and coworkers.   The panel (c) of Fig.~\ref{fig:wavy_roll_energy} shows the time series of the modes $W_{101}$, $W_{111}$, and $Z_{010}$ that illustrates their growth and subsequent breakdowns (critical bursting).  Note that the time-scales of oscillations for the modes $W_{111}$ and $Z_{010}$ are around 0.1, which is in  the same range as the theoretical time-scale derived above using the bifurcation analysis.

The above arguments strongly suggest that the origin of the wavy rolls or the oscillatory instabilities are intimately related to the purely imaginary pair of eigenvalues at CB2 and the limit cycles that originate from it.  

\section{Conclusion}\label{conclusion}

In conclusion, we explored various flow patterns of zero-P convection and performed a detailed bifurcation analysis near the onset using direct numerical simulation and a 27-mode low-dimensional model.   The low-dimensional model was constructed using the most energetic modes of DNS.  The results of the DNS and the low-dimensional model are in good agreement with each other.  Several new chaotic attractors and windows of periodic and quasiperiodic rolls  have been reported for the first time for zero-P convection.   The origin and dynamics of all the observed patterns have been explained successfully using the bifurcation diagrams.

The RBC system for $P=0$ is chaotic at the onset itself.  The stability analysis of the 27-mode model indicates  three codimension-2 bifurcation points that play critical roles in the dynamics of convection near the onset.   The chaotic attractors {\em Ch1}, {\em Ch2}, and {\em Ch3}, described earlier by \cite{Pal:EPL_2009}, and {\em Ch4} are all related to the bifurcation point CB3.   Beyond {\em Ch1} and {\em Ch4}, we observe SQOR, OASQ, ASQ, SQ etc., that are common to Pal \etal's 13-mode model.  The other codimension-2 bifurcation point, CB2, generates chaotic attractors {\em Ch5} and {\em Ch6}, and the subsequent periodic, quasiperiodic, and phase-locked convective states, which correspond to the wavy roll patterns of RBC observed earlier in experiments and simulations.      In addition,  we find that the frequency of the wavy rolls are connected to the imaginary eigenvalues of the stability matrix at the CB2 bifurcation point.  Thus, the bifurcation analysis presented in the paper provides useful insights into the origin of the wavy rolls of RBC.  

Interestingly, the bifurcation diagram of 30-mode model of \cite{mishra:EPl_2010} for $P=0.0002$ matches quite closely with Bif-A of our model.  This reinforces earlier observations that zero-P convection is a valid limit of low-P convection as $P \rightarrow 0$ \cite[]{thual:zeroP_1992}.   The extension of the present study to low-P convection in relation to wavy rolls will be very valuable for understanding the experimental and numerical findings near the onset.

\begin{acknowledgments}
We are thankful to Krishna Kumar and Pankaj K. Mishra for useful discussions. This work is supported by the Swarnajayanti fellowship grant to MKV by Department of Science and Technology, India.
\end{acknowledgments}

\end{document}